\documentstyle[12pt]{article}
\title{On the nature of superconductor pseudogaps}
\author{N.~Kristoffel\\
Institute of Physics, University of Tartu,\\ Riia 142,
51014 Tartu, Estonia\\
Institute of Theoretical Physics, University of Tartu,\\
T\"ahe 4, 51010 Tartu, Estonia}
\date{}
\begin{document}
\maketitle

\begin{abstract}
The physical origin of cuprate high-temperature superconductor pseudogaps
remains debatable. We point out that the indication of such excitation is
hidden in the usual expression for the quasiparticle energy. It can be
realized on a suitable multiband spectrum with an interband pairing channel.
The band components bearing the chemical potential manifest
superconducting gaps. A band with the Fermi energy outside creates a pseudogap
type excitation. The latter does not characterize the pairing strength.
On a doping-driven spectral arrangement the nature of low-energy
excitations changes with doping. The pseudogap appears as a precursor of
the corresponding superconducting gap on the doping scale. The
corresponding critical points on the phase diagram are determined by
the doping-driven overlap dynamics of the bare gapped electron spectrum.
\end{abstract}

%\pacs{74.20.-z; 74.72.-h}


%\maketitle

%\section{Introduction}
The pseudogap conception is connected with a peculiarity in the excitation
spectrum of cuprate high-temperature superconductors. A depletion of
low-energy excitation density around the Fermi level is observed in both the
superconducting and the normal state. Since this gap feature survives in a
remarkable extent at $T>T_c$ it has been designated as a pseudogap [1-4].

The pseudogap energy scale markedly exceeds the condensation energy,
especially in the underdoped region. With an extended doping the pseudogap
decreases and is quenched at a slight overdoping.
Further, the spectrum remains
determined by the superconducting gap(s). A magnetic counterpart of the
charge-channel pseudogap is also known in the spin-excitation spectrum.
The physical origin of the pseudogap remains a widely debated problem
with no consensus reached. Two major approaches to the nature of the
pseudogap have been elaborated: (i) the internal, and (ii) the extrinsic one.

The internal approach relates the pseudogap immediately with the pairing
strength, i.e. considers it as a precursor of the superconducting gap
on the energetic scale. Here belong the approaches with preformed
pairs (without any phase coherence at $T>T_c$), or the ones
based on superconducting order fluctuations [1-6].
Recent experimental data seem to prefer the extrinsic
scenario \cite{2,4}. Here the pseudogap source is reduced to bare
normal-state gaps of various origins [1-3,7-10].

In the extrinsic mechanisms a gapped (at least a two-band) system must
show the superconductivity besides the pseudogap. The two-gap behaviour
of cuprates has been revealed by recent spectroscopy data [11-13]. As a
minimum, it means the presence of a pseudogap plus a superconducting gap,
or two pseudogaps and a superconducting gap [14-16], on the same doping.
Indeed, in a gapped two-band system a spontaneous appearance of the
superconductivity is possible \cite{17}, if the condensation energy prevails
the bare gap. However, another highly-effective pairing channel can be
operative [17-19]. It consists in the pair-transfer between band
components. This mechanism, known already for a considerable time
\cite{20,21}, provides the simplest way to reach high transition temperatures
in a multiband system. Here the pairing can arise by repulsive interband
interaction, which operates in a considerable volume of the momentum space.
There has been a number of multiband approaches to cuprate superconductivity
(e.g. for review [18,22-26]). However the nature of coupled band
components has often remained unspecified or without justification.

In this letter, I emphasize that the pseudogap can appear naturally as a
minimal quasiparticle excitation energy in a multiband system with the
interband pairing. The simplest representative system will include two
gapped bands coupled by the pair transfer channel with the chemical
potential ($\mu$) intersecting only one of them.

In a two-band model of superconductivity the usual expression for the
quasiparticle energies
\begin{equation}
E_{\sigma}=[(\epsilon_{\alpha}-\mu)\sp 2+\Delta_{\alpha}\sp 2]\sp{1/2}
\end{equation}
holds \cite{21}. Here $\epsilon_{\alpha}$ are the band energies and
$\Delta_{\alpha}$ the superconducting gaps. For a band bearing the chemical
potential $E_{\sigma}$ is minimized at $\epsilon_{\alpha}=\mu$ and the
low-energy excitations manifest the superconducting gap. In the opposite
case
\begin{equation}
E_{\tau}(min)=[(\epsilon_{\tau}(e)-\mu )_m\sp 2+\Delta_{\tau}\sp 2]\sp{1/2}\; .
\end{equation}
In this expression the minimal value $(\epsilon_{\tau}(e)-\mu)_m$, with
$\epsilon_{\tau}(e)$ being the $\tau$-band edge, reflects the presence of a
normal state $\sigma$-$\tau$ gap ($\mu$ out of $\epsilon_{\tau}$). In the
normal state $E_{\tau}(min)$ survives and the excitations of this band correspond
to the pseudogap $\Delta_p=E_{\tau}(min)$. The changes in the band structure
and $\mu$ by doping can quench the pseudogap. Then the system will be
characterized by two superconducting gaps. The bare gap contribution to
the pseudogap energy can  markedly exceed the superconducting contribution.

Spectrally the smaller of the gaps, also in the presence of $\Delta_p$,
becomes manifested as an additive density inside the larger one. Our
approach to the pseudogap formation exposes it as a precursor of the
superconducting gap on the doping scale. It cannot be considered as a
measure of the condensation energy.

The following illustration concerns the cuprate superconductors. However, the
justification of the used model and a discussion of the results of its
application remain out of the scope of the present letter. We use this model
only to illustrate the natural appearance of the pseudogap on a nonrigid
bare gapped spectrum of a doped charge -- transfer insulator with the interband
pairing channel.

Cuprate superconductivity as such is stimulated by doping and the associated
characteristics depend strongly on doping. The structure of doped cuprates
has been found to be inhomogeneous on the nanoscale (stripes, tweed patterns,
granularity) with the associated electronic phase separation in the CuO$_2$
planes. A new distribution of doping-induced states appears in the
charge-transfer gap near the Fermi energy [26-28]. Various data indicate
the functioning of
itinerant and "defect''-type carriers in the basic physics of cuprate
superconductivity \cite{29}. Correspondingly in Refs. [18,19,30] a simple
model has been developed to describe such two-component scenario.
An idea that the hole doping creates not only the carriers
but prepares also the whole background with a new pairing channel for the
cuprate superconductivity has been elaborated. The hole-poor material can be
considered as remaining the source for the itinerant type band (of mainly
oxygen origin between the Cu dominated Hubbard components) and the
part of distorted material bearing the doped holes as creating defect bands.
The bare gaps between these subsystems, quenched by a progressive doping, have
been supposed to be the origin for the pseudogap behaviour. The interband
pairing between the itinerant and defect subsystem is postulated to be
the leading pairing mechanism. The corresponding theoretical formulation
can be followed by Refs. [19,30].

The band arrangements of the model \cite{30} are schematized in Fig.1.
The case a) corresponds to a heavily underdoped region. The $\alpha$ and
$\beta$ bands represent the "hot'' ($\pi ,0$) and "cold''
($\frac{\pi}{2},\frac{\pi}{2})$ regions of the momentum space and they belong
to the defect subsystem. Experimentally it is well known that doping brings
the defect states to merge with the basic itinerant band.An extended
doping shifts correspondingly the bottoms of these bands down in energy,
leading to the bands overlap. At moderate dopings the cold quasiparticles
are metallic while the hot ones remain insulating.
In the case b) the bare $\beta$-$\gamma$ gap
($\gamma$ designates the itinerant band) is quenched and $T_c$ grows
until the optimal doping is reached, as shown on Fig.1c. The optimal
doping corresponds to the overlap of all the band components being
intersected by $\mu$. Further doping
deteriorates the conditions for the leading $(\alpha ,\beta )-\gamma$ pairing.
The nongapped mixed spectrum reflects the restoring of the normal Fermi
liquid behaviour on overdoping.

The calculated gaps of the model are illustrated on the whole hole doping
scale ($p$) in Fig.2. The model contains two superconducting gaps $\Delta_{\gamma}$
and $\Delta_{\alpha}=\Delta_{\beta}$ (taken for simplicity) and two pseudogaps
$\Delta_{p\alpha}$ (the larger one) and $\Delta_{p\gamma}$.
For the visual purpose only the gap complex connected to the
"$\Delta_{p\alpha}$ driven phase'' is shown together with $T_c$.
In the case a) one
expects the observing of two pseudogaps (like in [14-16]).
The smaller pseudogap $\Delta_{p\gamma}$ is lost
when the $\beta$-$\gamma$ overlap is reached. In the case of missing
$\beta$-$\gamma$ bare gap there will be only one pseudogap. However, the
participation of $\beta$ subsystem is essential for increasing $T_c$ as
the partner in the interband pairing. In the case of larger
dopings $\Delta_{p\alpha}$ and $\Delta_{\gamma}$ reside until the large
pseudogap is quenched for the bands arrangement in Fig.1c. The overdoped
region is represented by two superconducting gaps $\Delta_{\alpha ,\gamma}$.
The itinerant and $\alpha$ band excitations represent the "hot'' spectrum.
The "cold'' part of the defect subsystem spectrum becomes empty for the
d-wave ordering.

The crossing of the large pseudogap $\Delta_{p\alpha}$, corresponding to
the spectral "hump'' \cite{31,32}, and of the larger superconducting
gap $\Delta_{\gamma}$ occurs close to the optimal doping (cf. the experiment
in \cite{31}). These gaps belong to different subsystems with noncompeting
order parameters vanishing at $T_c$ simultaneously.
The manifestation of a  superconducting gap at a given
doping can be substituted by the appearance \cite{32} of the normal state
gap  for $T>T_c$. It means that at low temperatures
a pseudogap may not manifest itself on dopings, where it will be found in
the normal state (cf. \cite{33}). The experimental data cannot be
interpreted \cite{32} as a transformation of a pseudogap into a
superconducting gap of the same subsystem on the energetic (pairing strength
\cite{5}) scale. The pseudogaps transform smoothly into superconducting
gaps with an extended doping on the doping-scale, as illustrated in Fig.2 and
as known experimentally [14-16,33,34]. Note that the nature of the low-energy
quasiparticle excitations changes with doping. The doping-driven spectral
overlap appears this way as a novel source of critical doping concentrations
on the phase diagram. In the normal state insulator to metal transitions
are expected at these points, cf. \cite{35}.

The transition temperature (Fig.2) and the superfluid density \cite{36}
show the usual bell-like behaviour with doping. The bare normal state gaps
 do not manifest themselves as fermionic gaps in the superfluid density because of the
interband nature of the doping. An argument against the "extrinsic'' nature
of the pseudogap \cite{9} falls out. Various observed relations between the
pseudo-, superconducting and normal state gaps on the cuprate phase
diagram can be explained in the described way.

The author is grateful to P.Rubin and T.\"Ord for a long-running
collaboration.

This work was supported by the Estonian Science Foundation Grant Nos 4961
and 6540. \\ \\

\newpage

Figure captions\\ \\

Figure 1. The band arrangements evolution with hole doping ($p$): \\
a) heavy underdoping; b) extended doping; c) optimal doping.
$\alpha$ and $\beta$ designate the defect system subbands (normalized
 to $p/2$); the itinerant band $\gamma$ (only its upper part is shown)
 is normalized to $1-p$. The horizontal sections of the bands reflect
the densities of states.\\

Figure 2. Gaps and the transition temperature of a "typical''
cuprate on the hole doping scale. \\
Curve 1 -- the large pseudogap $\Delta_{p\alpha}$; curve 2 -- the
itinerant subsystem
superconducting gap; curve 3 -- the defect subsystem superconducting gap;
curve 4 -- $T_c$.

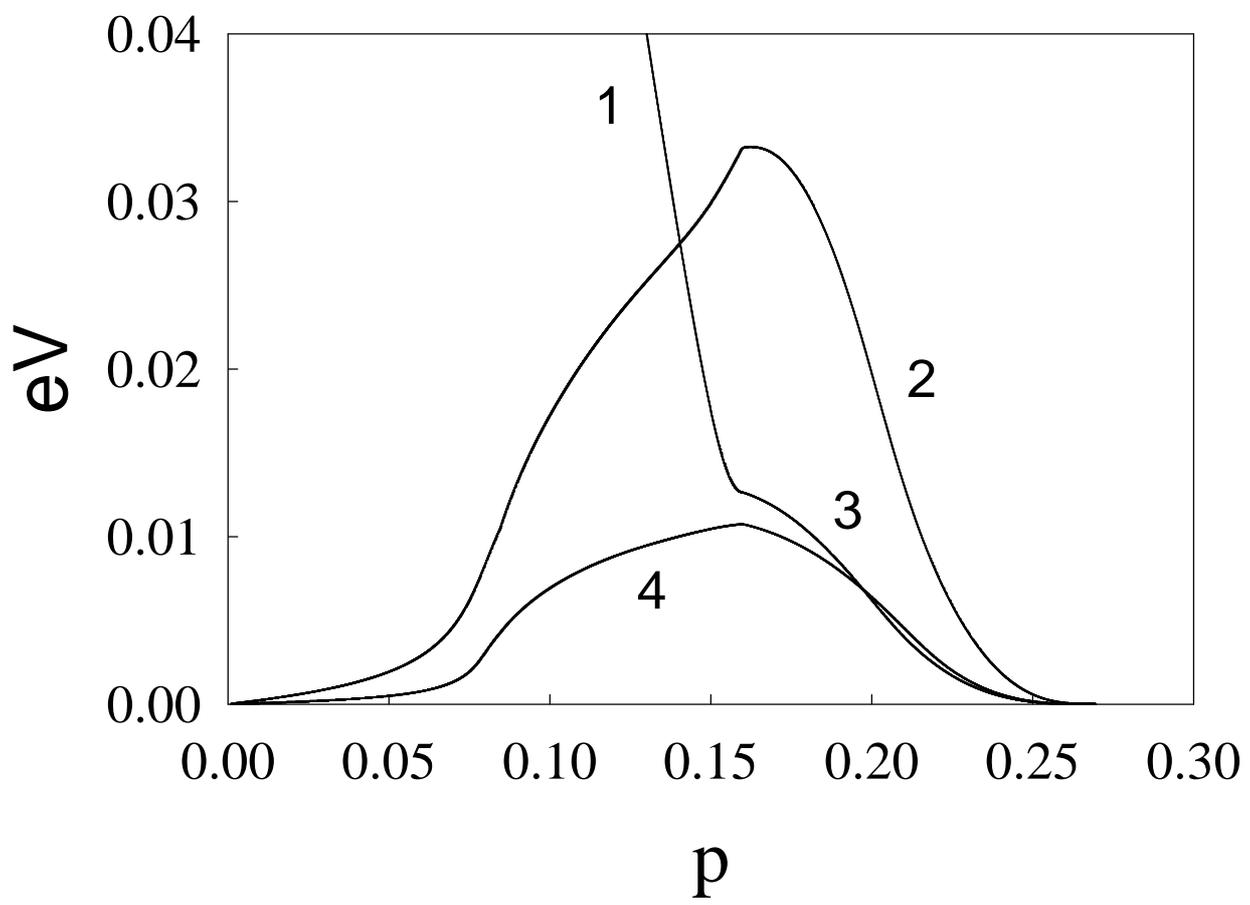
\end{document}